\documentclass{pasj01}

\def\commenta{$^*$}
\def\commentb{$^\dagger$}
\def\commentc{$^\ddagger$}
\def\commentd{$^\S$}
\def\commente{$^\|$}
\def\commentf{$^\#$}

\newcounter{author}
\def\authorcount#1#2{\refstepcounter{author}\label{#1}
                     \altaffiltext{\ref{#1}}{#2}}

\begin{document}
\SetRunningHead{K. Isogai et al.}{CR Boo}

\Received{201X/XX/XX}
\Accepted{201X/XX/XX}

\title{Superoutburst of CR Bootis: Estimation of Mass Ratio of a typical AM CVn star by Stage A Superhumps}

\author{Keisuke~\textsc{Isogai},\altaffilmark{\ref{affil:Kyoto}*}
Taichi~\textsc{Kato},\altaffilmark{\ref{affil:Kyoto}}
Tomohito~\textsc{Ohshima},\altaffilmark{\ref{affil:Nishiha}}
Kiyoshi~\textsc{Kasai},\altaffilmark{\ref{affil:Kai}}
Arto~\textsc{Oksanen},\altaffilmark{\ref{affil:Nyrola}}
        Kazunari~\textsc{Masumoto},\altaffilmark{\ref{affil:OKU}}
        Daiki~\textsc{Fukushima},\altaffilmark{\ref{affil:OKU}}
        Kazuki~\textsc{Maeda},\altaffilmark{\ref{affil:OKU}}
        Miho~\textsc{Kawabata},\altaffilmark{\ref{affil:OKU}}
        Risa~\textsc{Matsuda},\altaffilmark{\ref{affil:OKU}}
        Naoto~\textsc{Kojiguchi},\altaffilmark{\ref{affil:OKU}}
        Yuki~\textsc{Sugiura},\altaffilmark{\ref{affil:OKU}}
        Nao~\textsc{Takeda},\altaffilmark{\ref{affil:OKU}}
        Katsura~\textsc{Matsumoto},\altaffilmark{\ref{affil:OKU}}
Hiroshi~\textsc{Itoh},\altaffilmark{\ref{affil:Ioh}}
	Elena~P.~\textsc{Pavlenko},\altaffilmark{\ref{affil:CrAO}}
	Kirill~\textsc{Antonyuk},\altaffilmark{\ref{affil:CrAO}}
	Oksana~\textsc{Antonyuk},\altaffilmark{\ref{affil:CrAO}}
	Nikolai~\textsc{Pit},\altaffilmark{\ref{affil:CrAO}}
	Aleksei~\textsc{Sosnovskij},\altaffilmark{\ref{affil:CrAO}}
	Alex~\textsc{Baklanov},\altaffilmark{\ref{affil:CrAO}}
	Julia~\textsc{Babina},\altaffilmark{\ref{affil:CrAO}}
	Aleksandr~\textsc{Sklyanov},\altaffilmark{\ref{affil:Kazan}}
Seiichiro~\textsc{Kiyota},\altaffilmark{\ref{affil:Kis}}
Franz-Josef~\textsc{Hambsch},\altaffilmark{\ref{affil:GEOS}}$^,$\altaffilmark{\ref{affil:BAV}}$^,$\altaffilmark{\ref{affil:Hambsch}}
Colin~\textsc{Littlefield},\altaffilmark{\ref{affil:LCO}}
Yutaka~\textsc{Maeda},\altaffilmark{\ref{affil:Mdy}}
Lewis~M.~\textsc{Cook},\altaffilmark{\ref{affil:LewCook}}
Gianluca~\textsc{Masi},\altaffilmark{\ref{affil:Masi}}
Pavol~A.~\textsc{Dubovsky},\altaffilmark{\ref{affil:Dubovsky}}
Rudolf~\textsc{Nov\'ak},\altaffilmark{\ref{affil:Novak}}
Shawn \textsc{Dvorak}, \altaffilmark{\ref{affil:DKS}}
Akira~\textsc{Imada},\altaffilmark{\ref{affil:Kwasan}}
Daisaku~\textsc{Nogami}\altaffilmark{\ref{affil:Kyoto}}
}

\authorcount{affil:Kyoto}{
     Department of Astronomy, Kyoto University, Kyoto 606-8502, Japan}
\email{$^*$isogai@kusastro.kyoto-u.ac.jp}

\authorcount{affil:Nishiha}{
        Nishi-Harima Astronomical Observatory, University of Hyogo, Japan}

\authorcount{affil:Kai}{
     Baselstrasse 133D, CH-4132 Muttenz, Switzerland}

\authorcount{affil:Nyrola}{
     Hankasalmi observatory, Jyvaskylan Sirius ry, Verkkoniementie
     30, 40950 Muurame, Finland}

\authorcount{affil:OKU}{
Osaka Kyoiku University, 4-698-1 Asahigaoka, Kashiwara, Osaka 582-8582, Japan}

\authorcount{affil:Ioh}{
     Variable Star Observers League in Japan (VSOLJ),
     1001-105 Nishiterakata, Hachioji, Tokyo 192-0153, Japan}

\authorcount{affil:CrAO}{
     Crimean Astrophysical Observatory, p/o Nauchny, 298409,
     Republic of Crimea}

\authorcount{affil:Kazan}{
     Kazan Federal University, Kremlevskaya str., 18, Kazan, 420008, Russia}

\authorcount{affil:Kis}{
     VSOLJ, 7-1 Kitahatsutomi, Kamagaya, Chiba 273-0126, Japan}

\authorcount{affil:GEOS}{
     Groupe Europ\'een d'Observations Stellaires (GEOS),
     23 Parc de Levesville, 28300 Bailleau l'Ev\^eque, France}

\authorcount{affil:BAV}{
     Bundesdeutsche Arbeitsgemeinschaft f\"ur Ver\"anderliche Sterne
     (BAV), Munsterdamm 90, 12169 Berlin, Germany}

\authorcount{affil:Hambsch}{
     Vereniging Voor Sterrenkunde (VVS), Oude Bleken 12, 2400 Mol, Belgium}

\authorcount{affil:LCO}{
     Department of Physics, University of Notre Dame, 
     225 Nieuwland Science Hall, Notre Dame, Indiana 46556, USA}

\authorcount{affil:Mdy}{
     Kaminishiyamamachi 12-14, Nagasaki, Nagasaki 850-0006, Japan}

\authorcount{affil:LewCook}{
     Center for Backyard Astrophysics Concord, 1730 Helix Ct. Concord,
     California 94518, USA}

\authorcount{affil:Masi}{
     The Virtual Telescope Project, Via Madonna del Loco 47, 03023
     Ceccano (FR), Italy}

\authorcount{affil:Dubovsky}{
     Vihorlat Observatory, Mierova 4, 06601 Humenne, Slovakia}

\authorcount{affil:Novak}{
     Research Centre for Toxic Compounds in the Environment, Faculty of 
     Science, Masaryk University, Kamenice 3, 625 00 Brno, Czech Republic}

\authorcount{affil:DKS}{
  Rolling Hills Observatory, 1643 Nightfall Drive, Clermont, Florida 34711, USA}

\authorcount{affil:Kwasan}{
        Kwasan-Hida Observatory, University of Kyoto, Japan}



\maketitle

\begin{abstract}
We report on two superoutbursts of the AM CVn-type object CR Boo in 2014 April--March and 2015 May--June.
A precursor outburst acompanied both of these superoutbursts.
During the rising branch of the main superoutburst in 2014, 
we detected growing superhumps (stage A superhumps) whose period was $0.017669(24)$ d.
Assuming that this period reflects the dynamical precession rate at the radius of the 3:1 resonance,
we could estimate the mass ratio ($q=M_2/M_1$) of 0.101(4)
by using the stage A superhump period and the orbital one of 0.0170290(6) d.
This mass ratio is consistent with that expected by the theoretical evolutionary model of AM CVn-type objects.
The detection of precursor outbursts and stage A superhumps is the second case in AM CVn-type objects.
There are two interpretations of the outbursts of AM CVn-type objects.
One is a dwarf nova (DN) outbursts analogy, which is caused by thermal and tidal instabilities.
Another is the VY Scl-type variation, which is caused by the variation of the mass-transfer rate of the secondary.
This detection of the superhump variations strongly suggests the former interpretation.
\end{abstract}

\section{Introduction}\label{sec:int}
AM CVn-type objects are a rare subclass of cataclysmic variables (CVs) 
which are close binary systems composed of a white dwarf (WD) primary 
and a mass-transferring secondary star.
They are characterized by absence of hydrogen lines in spectra and 
their ultra-short orbital periods of ${\sim}$5--65 min.
These features suggest that the secondary is a helium WD. 
AM CVn-type objects are expected as one of the most promising sources of 
low-frequency gravitational wave (GW) radiation,
which are detectable by Evolved Laser Interferometer Space Antenna (eLISA) missions \citep{nel13elisa}.
Moreover, some of them are believed to be the progenitor of some of the type Ia supernovae \citep{sol05amcvnSN}.
It is thus important to inspect the evolutionary model of AM CVn-type objects.
(for reviews, see \cite{nel05amcvnreview,sol10amcvnreview}).

Three evolutionary channels (WD, He-star, and hydrogen-star channels) have been proposed to form AM CVn-type objects.
The former two channels experience two common envelope (CE) events.
The channels can be distinguished whether the remnant core of the donor is 
a degenerate WD or non-degenerate He-star.
GW radiation makes the remnant cores of main-sequence (MS) stars close,
and the semi-detached or detached binaries are formed.
The third channel experiences one CE event and evolves in the same way as hydrogen-rich CVs.
However, the binaries enable more compact orbits because of the hydrogen deficiency of the donor.
The population synthesis, which depends on the birthrate in these three channels,
has been discussed, but is still poorly understood.
In order to investigate the population, the detailed observations are essential.
By measuring the donor mass and the orbital period,
we will be able to clarify whether the donor is a semi-degenerate or fully-degenerate
by the theoretical equations listed below.
Moreover, in each channel, the binary system once or twice experiences common envelope (CE) phases.
Therefore the study of AM CVn-type objects may help elucidate yet poorly understood CE phenomena.
For more details, see \citet{del07amcvn} and \citet{sol10amcvnreview}.

\citet{fau72amcvn}, \citet{zap69massradius} and \citet{sav86periodminimum} derived the following equations
with requirements for the Roche-lobe overflow and He-star secondary, respectively:
\begin{equation}\label{eq:1}
P_{\rm orb} ({\rm h}) = 8.75 ( M_2/R_2^3 )^{-1/2}{\rm ,}
\end{equation}
\begin{equation}\label{eq:2}
R_2 = 0.0155M_2^{-0.212} \ \ \ {\rm for\ the \ fully\mathchar`-degenerate\ secondary,}
\end{equation}
\begin{equation}\label{eq:3}
R_2 = 0.029M_2^{-0.19} \ \ \ {\rm for\ the \ semi\mathchar`-degenerate\ secondary,}
\end{equation}
where $P_{\rm orb}$ is the orbital period, $M_2$ is the secondary mass and $R_2$ is 
the secondary radius in solar units, respectively.
Equations (\ref{eq:1}), (\ref{eq:2}), (\ref{eq:3}) yield
\begin{equation}\label{eq:fulldege}
M_2 = 0.0069P_{\rm orb}^{-1.22} \ \ \ \ \ {\rm for\ the \ fully\mathchar`-degenerate\ secondary,}
\end{equation}
\begin{equation}\label{eq:semidege}
M_2 = 0.018P_{\rm orb}^{-1.27} \ \ \ \ \ {\rm for\ the \ semi\mathchar`-degenerate\ secondary,}
\end{equation}
Based on equations (\ref{eq:fulldege}), (\ref{eq:semidege}),
we can draw the theoretical evolutionary tracks which represent
the mass-radius relations of the binary systems with fully and semi-degenerate secondaries.
If we obtain the mass ratios of AM CVn-type objects, 
we can test the suitability of the evolutionary model.

Outburst phenomena in AM CVn-type objects are widely known.
There is a theoretical model of outbursts in AM CVn-type objects \citep{tsu97amcvn}.
Most of the outbursting AM CVn-type objects, 
whose orbital periods are $\sim$1300--2500 s, show superoutbursts and superhumps,
because they have sufficiently low mass ratios and 
the outer edges of the disks can reach the 3:1 resonance radius ($r_{\rm{3:1}}$)
which causes the tidal instability \citep{whi88tidal,osa89suuma,lub91SHa,lub91SHb,hir90SHexcess}.
In hydrogen-rich CVs, three stages of superhumps were discovered by \citet{Pdot}.
\citet{kat13qfromstageA} partially succeeded in interpreting these stages of superhumps.
The stage A is the growing stage of superhumps and 
its period is thought to reflect the dynamical precession rate at $r_{\rm{3:1}}$.
The pressure effect makes the superhump period suddenly shorter at the start of the stage B.
The superhump period then become larger gradually.
The stage C shows a constant shorter period than in the stage B, but the origin of the stage C is still unclear.
On the basis of this interpretation, we can estimate the mass ratio by measurements of the stage A superhump period and the orbital one.
In this paper, we applied this method to CR Boo.

\section{CR Bootis}
CR Boo is one of best-known AM CVn-type objects, 
which shows SU UMa-type outburst phenomena \citep{kat99erumareview}.
No one, however, has ever performed time-resolved photometric observations through the superoutburst in detail.
This object, called PG 1346+082, was discovered by the Palomar Green survey \citep{gre86pgcatalog}.
\citet{woo87crboo} reported on the photometric and spectroscopic variations,
and it was confirmed that this object is an interacting double white dwarf binary.
The orbital period is 0.0170290(6) d \citep{pro97crboo}.
Although the period variation of superhump which is
similar to hydrogen-rich CVs was reported in \citep{Pdot4},
we could see the growing stage of superhumps for the first time in our present observations.

CR Boo has two distinct states: (1) fainter quiescence with regular superoutbursts state, 
which corresponds to the ``ER UMa-like'' state, 
and (2) brighter quiescence with frequent outbursts state \citep{hon13crboo,Pdot4}.
During our observations, CR Boo was in the former state.
ER UMa-type is a subclass of SU UMa-type CVs and
is characterized by an extreme high outburst frequency
and short supercycles ($\sim$19--48 d) \citep{kat95eruma},
which is considered to be caused by high mass-transfer rates from the donor star \citep{osa95eruma}.
In this state, CR Boo shows a $\sim46$ d supercycle \citep{kat99erumareview}.
We can predict the date of next superoutburst by using this supercycle.
We performed two observation campaigns.

\section{Observation and Analysis}\label{sec:obs}
 Our time-resolved photometric campaigns were carried out 
in 2014 April--May and 2015 May--June
by the Variable Star Network (VSNET) collaborations \citep{VSNET}.
The data were acquired with 20--40cm class telescopes.
The logs of photometric observations are summarized in table $1$ and $2$ 
(all tables are reported only in the electronic edition).
The data were observed by $V$-band and unfiltered CCD photometry.
In the normal outbursting CVs, the magnitude of unfiltered CCD photometry is 
close to $V$-band magnitude, and we treated them in the same one.
The time of the observations were corrected to barycentric Julian date (BJD).
We corrected the zero-point of each observer's data 
by adding a constant for each observer.

We used the phase dispersion minimization (PDM) method \citep{PDM}
for analyzing the periodic variations.
The $1\sigma$ errors by this method were determined by the methods described in \citet{fer89error} and \citet{Pdot2}.
Before performing period analyses, we subtracted the global trend of
the light curve by subtracting smoothed light curve 
obtained by locally-weighted polynomial regression (LOWESS, \cite{LOWESS}).
We used O-C diagrams which are sensitive to subtle variations of
the superhump period (see \cite{ste05oc}).
We can identify stage transitions of superhumps by detecting kinks in the $O-C$ diagrams.
The times of superhump maxima, which are used to draw the $O-C$ diagram, were determined 
by the same method as described in \citet{Pdot}.

\section{Result}\label{sec:result}
\subsection{Overall Light Curve}\label{sec:lc}
\begin{figure}
\begin{center}
    \FigureFile(80mm,50mm){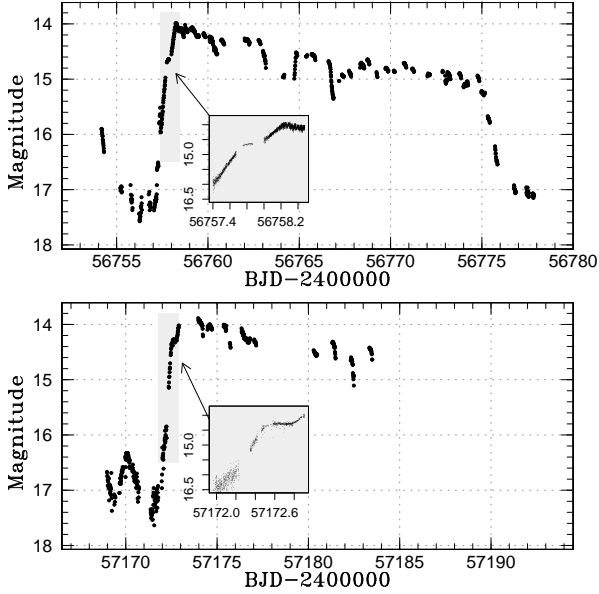}
\end{center}
\caption{Overall light curves of CR Boo in 2014 April--May (Upper panel) and in 2015 May--June (Lower panel).
The observations were binned to 0.01 d.
Insets: the enlarged light curves, correspond to the shaded areas, around the precursor outbursts indicated by the arrows.}
\label{fig:lc}
\end{figure}

The upper panel of figure \ref{fig:lc} shows the overall light curve of CR Boo in 2014 April--May.
As shown in this figure, the superoutburst lasted for 17 days.
The magnitude in quiescence was $V \sim17.2$ and 
the brightness rapidly increased to $V\sim14$ on BJD 2456757.
We can see a precursor outburst as a form of a ``kink'' in the light curve during BJD 2456757--2456758
(see the inset in the upper panel of figure \ref{fig:lc}).
Following the precursor outburst, the main superoutburst began.
During the plateau phase, unlike typical superoutbursts in hydrogen-rich CVs, 
we can see the rapid fading and brightening, which we call a ``dip'', with small amplitudes of $\sim$0.5--1.0 mag.
The lower panel of figure \ref{fig:lc} shows the overall light curves of CR Boo in 2015 May--June.
The profile of the outburst was similar to the one in 2014, including a precursor outburst  BJD 2457172-2457172.8
(see the inset in the lower panel of figure \ref{fig:lc}).

\subsection{Superhumps}\label{sec:superhump}
\begin{figure}
\begin{center}
   \FigureFile(80mm,50mm){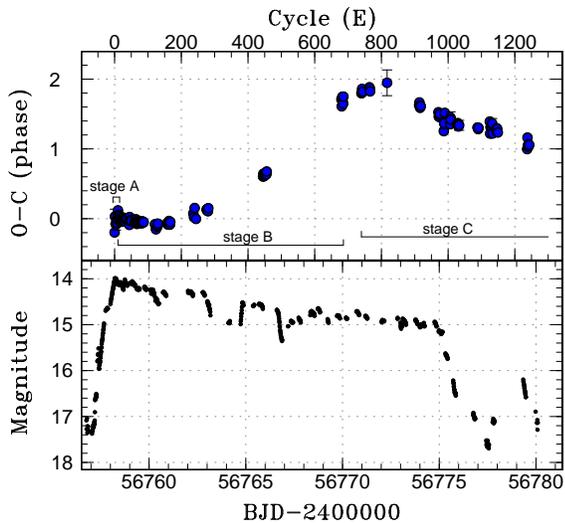}
\end{center}
\caption{(Upper) $O-C$ diagram of CR Boo during the 2014 April superoutburst.
An ephemeris of Max(BJD) $=2456758.0111 + 0.0172414 E$ was used to draw this figure.
The vertical axis indicates a phase (1 phase = 0.0172414 d).
The intervals stage A--C represents superhump stages.
  (Lower) Light curve. The observations were binned to 0.01 d.
The horizontal axis in units of BJD and cycle number is common to these three panels.
The light curve shows two dips around BJD 2456765 and 2456767.
Judging from the O-C diagram, the stage B superhumps continued through the dips.
However, the presence of dips suggests that 
the state may have been different from stage B superhumps in hydrogen-rich CVs.} 
\label{fig:oc14}
\end{figure}

\begin{figure}
\begin{center}
    \FigureFile(80mm,50mm){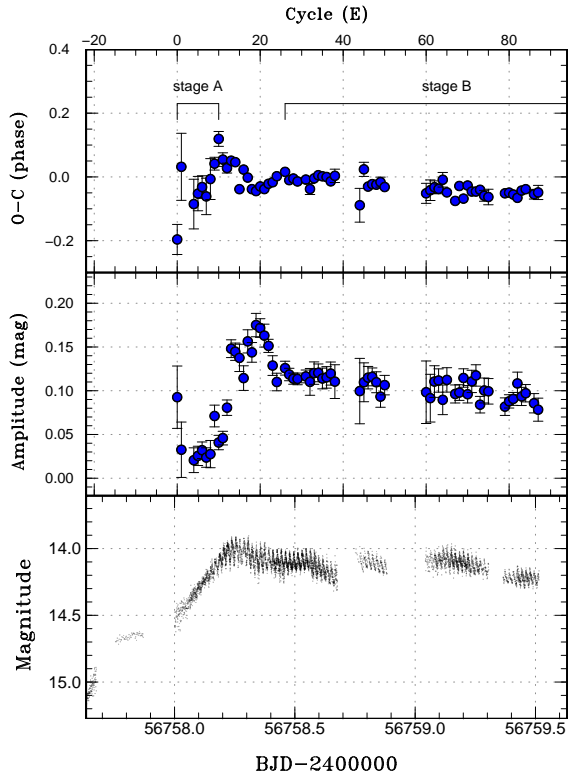}
\end{center}
\caption{Enlarged figure \ref{fig:oc14} around the stage A-B transition.
(Upper) $O-C$ diagram of CR Boo during the 2014 April superoutburst.
An ephemeris of Max(BJD) $=2456758.0111 + 0.0172414 E$ was used to draw this figure.
The vertical axis indicates a phase (1 phase = 0.0172414 d).
The intervals stage A--B represents superhump stages.
  (Middle) Amplitudes of superhumps.
  (Lower) Light curve.
The horizontal axis in units of BJD and cycle number is common to these three panels.} 
\label{fig:oc14AB}
\end{figure}

\begin{figure*}
\begin{center}
    \FigureFile(120mm,75mm){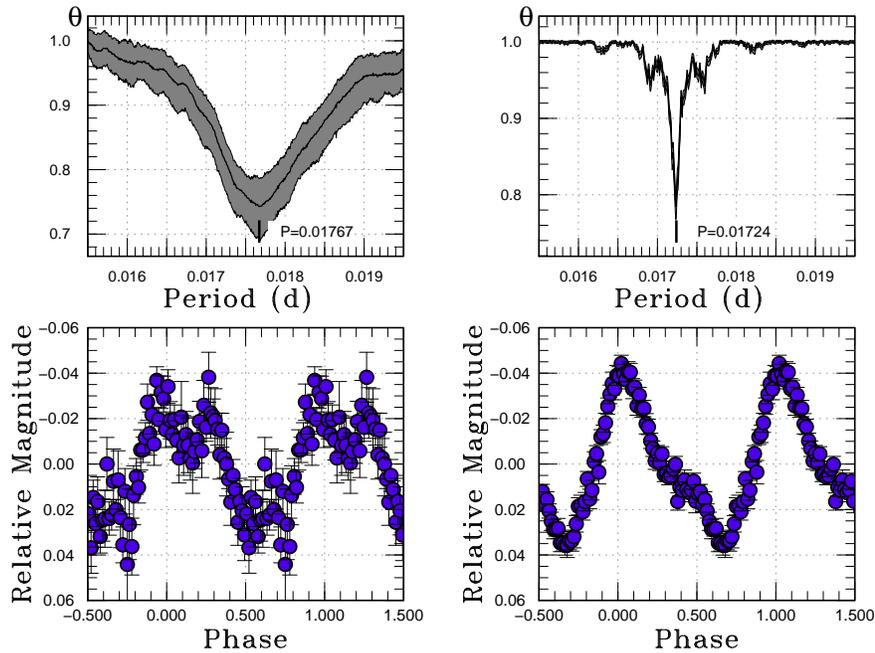}
\end{center}
\caption{Results of the period analyses of superhumps of the 2014 outburst in BJD 2456758.0--5456758.195 (Left) 
and in BJD 2456758.46--2456770 (Right) .
(Upper): $\theta$ diagram of our PDM analysis of superhumps.
The area of gray scale means $1 \sigma$ errors.
(Lower): Phase-averaged profiles of superhumps.
The left and right panels correspond to the period analyses 
of the stage A and the stage B superhumps, respectively.} 
\label{fig:pdm14}
\end{figure*}

Figure \ref{fig:oc14} shows the overall $O-C$ diagram (upper panel), and the light curve (lower panel) in the 2014 outburst. 
A period of 0.0172414 d, which is the stage B superhump period in 2014 we derived (see below), was used for calculating the O-C values.
The times of superhump maxima are listed in table 3.
As mentioned in section \ref{sec:int}, the superhumps can be classified into three stages 
by the features of the variations of the periods and amplitudes.
The transition of the tendency of the $O-C$ diagram around $E \sim 10$ and 700 
indicates the transition of the stage A--B and B--C, respectively.
Figure \ref{fig:oc14AB} is the enlarged figure \ref{fig:oc14} around the stage A--B transition,
and shows $O-C$ diagram (upper panel), the amplitude of superhumps (middle panel), and the light curve (lower panel). 
We can see that the superhumps started to develop 
during the rapidly rising phase (around BJD 2456758) following the precursor (lower panel).
As can be seen in this $O-C$ diagram, a kink can be seen at $E {\sim} 10$, 
corresponding to the maximum amplitude of the superhumps and the peak of the superoutburst. 
These features suggest that the maxima for $E {\sim}$0--10 can be regarded as the stage A of superhumps.
During $E = $11--26, the tendency of the $O-C$ diagram is unstable.
A transition between stage A and B thus occurred in this phase.
The superhump period gradually increased through the stage B ($E = $27--686) 
and the superhump amplitude was continuously decreased.
As shown in upper panels of figure \ref{fig:pdm14}, 
the estimated mean period of the stage A in BJD 2456758.0--2456758.195, 
the stage B in BJD 5456758.46--2456770.0 and the stage C in BJD 2456770.7--2456779.5 are
0.017669(24), 0.017241(18) and 0.017218(5) d, respectively.
The lower panels in figure \ref{fig:pdm14} are the phase-averaged profiles of the stage A and the stage B.
The profiles have a double-wave shape in the stage A and single-wave shape in the stage B, 
which match as the well-known profiles of superhumps (e.g. \cite{Pdot7}).

\begin{figure}
\begin{center}
   \FigureFile(80mm,50mm){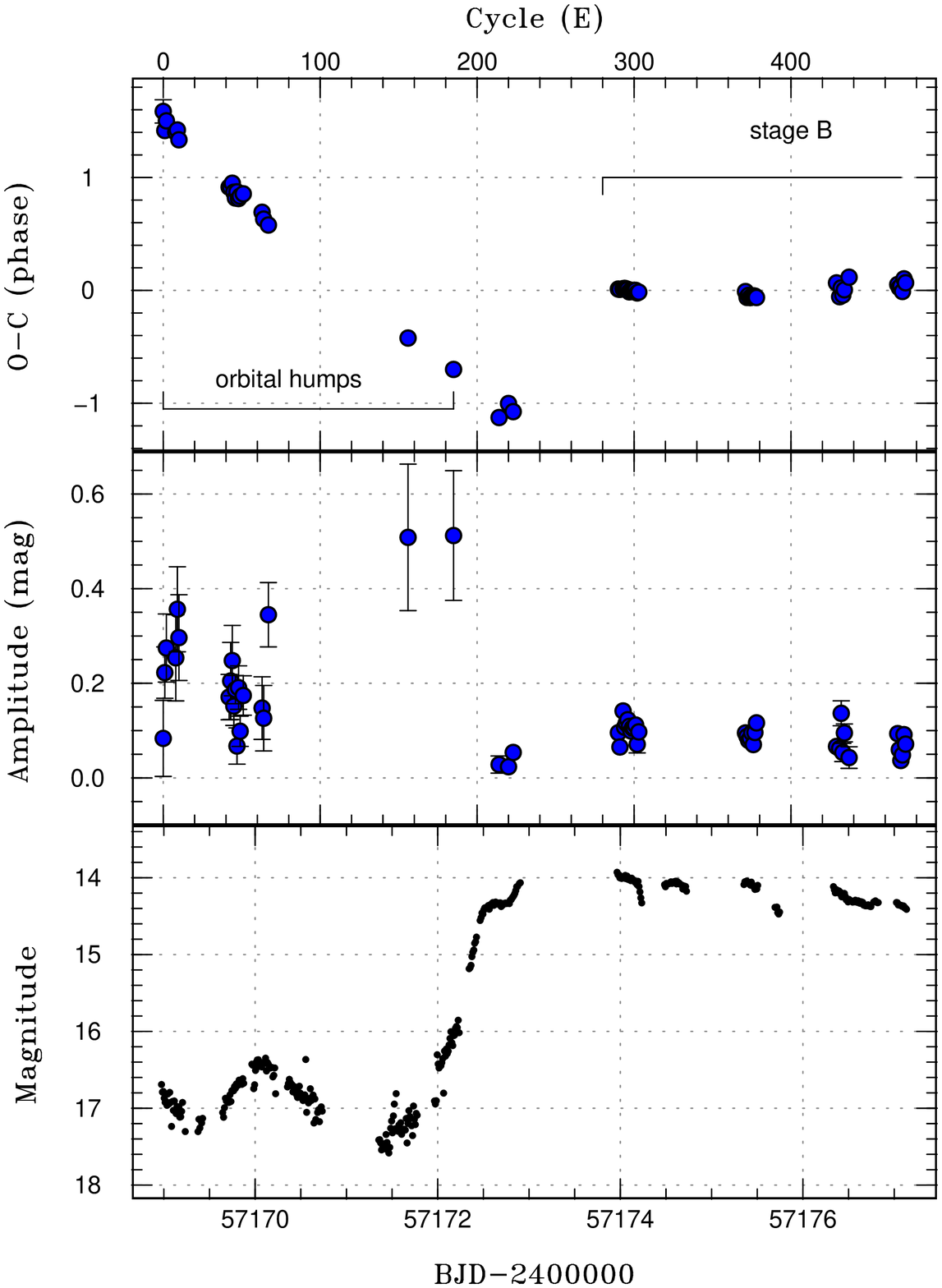}
\end{center}
\caption{(Upper) $O-C$ diagram of CR Boo during the 2015 May superoutburst.
An ephemeris of Max(BJD) $=2457168.99251 + 0.0172414E$ was used to draw this figure.
The vertical axis indicates a phase (1 phase $=0.0172414$ d).
The $O-C$ diagram in $E > 290$ is similar to stage B in the 2014 outburst.
The humps during $E =$ 0--185 appears to be the orbital ones because of its value.
Although we cannot see clear stage A superhumps.
  (Middle) Amplitudes of superhumps.
  (Lower) Light curve. The observations were binned to 0.01 d.
The horizontal axis in units of BJD and cycle number is common to these three panels.} 
\label{fig:oc15}
\end{figure}

Figure \ref{fig:oc15} shows the $O-C$ diagram of the 2015 outburst.
The times of superhump or orbital hump maxima are listed in table 4.
Unfortunately, the sparse data prevented us from detailed analyses of the stage A. 
The stage B superhumps ($E > 290$) with a period of 0.017237(48) d
are, however, consistent with those in the 2014 outburst.
The humps during $E =$ 0--185 with a period of 0.017031(6) d appears to be the orbital ones,
since their period is stably 0.017031(6) in good accordance with the previously measured one \citep{pro97crboo}.
During the precursor outburst, the $O-C$ diagrams showed increasing tendency ($E =$ 214--223).
Therefore we analyzed the period in BJD 2457172.55--2457173.00
where stage A superhumps were expected.
We estimated the period of 0.017591(20) d with the PDM method
which is in agreement with one in the 2014 within errors.
Thus this phase appears to be identified as the stage A.
This period, however, is not very reliable since the observation
contained only three maxima and some contamination of
the orbital variation may have been present.
Thus, we used the period of the stage A in 2015 for just a reference.

\section{Discussion}\label{sec:discussion}
\subsection{Evolution of Superhumps During the Superoutburst}
In the 2014 outburst, we succeeded in detecting the evolution of superhumps.
This is the first observation of the stage A superhumps in this object
and the second observation in all AM CVn-type objects \citep{kat14j0902}.
This fact suggests that we can apply the new method of mass-ratio estimation,
which uses the stage A superhump period and the orbital period,
not only for hydrogen-rich CVs but also for AM CVn-type objects.
However, the stage A in the 2014 outburst was observed for a cycle count of only 10
and the duration which corresponds to the phase 
from the precursor to the maximum of the superoutburst is $\sim0.5$ d.
Thus the detection of stage A in this object is difficult.
After stage A, the $O-C$ diagram for the 2014 outburst shows an unstable period during $E=$ 10--26.
We therefore considered that the stage A--B transitions of superhumps occurred here.

The lower panel of figure \ref{fig:pdm14} shows the profile of superhumps in the stages A and B.
The transition from double-wave modulations to single-wave ones
between the stage A and B is often seen in hydrogen-rich CVs \citep{Pdot7}.
The period derivative $P_{\rm{dot}}=\dot{P}/P$ of the stage B was 
$1.01(2) \times 10^{-5}$ in 2014 and $1.76(37) \times 10^{-5}$ in 2015.
Although the stage B in 2014 lasted at least for 660 cycles which is extremely long,
the duration of the stage B in 2012 March was about 300 cycles \citep{Pdot4}.
The presence of dips suggests that 
the state may have been different from stage B superhumps in hydrogen-rich CVs.
Thus we assumed that the stage B in 2014 is $E=$ 27--281.
The re-estimated $P_{\rm{dot}}(2014)$ was $1.76(12) \times 10^{-5}$ and
consistent with  $P_{\rm{dot}}(2012)=2.0(0.2) \times 10^{-5}$ \citep{Pdot4} and $P_{\rm{dot}}(2015)$.
This result indicates two possibilities.
The first possibility is that the stage B is not $E=$ 27--686 but 27--281.
The second possibility is that $P_{\rm dot}$ depends on the stage B duration.
Because $P_{\rm{dot}}$ is estimated on the assumption 
that the variation of stage B is expressed by a quadratic curve,
we cannot use a parabolic curve for the fitting, if the assumption is wrong.

\subsection{Outburst Behavior of CR Boo}
\citet{war95amcvn} interpreted and \citet{war15amcvnmemsai} examined the typical outbursting AM CVn-type objects,
whose orbital periods are 1300--2500 s, as VY Scl-type,
which is characterized by the unstable mass-transfer rate.
Moreover, \citet{war15amcvnmemsai} stated that 
``the variety of supercycle and the light curves do not convincingly support the interpretation as DN outbursts''.
However, the above superhump behavior clearly demonstrates that
these superoutbursts are DN outbursts analogues,
and the variations of the superhump period cannot be explained by
the luminosity variation of VY Scl-type.

During the plateau phase, we can see two dips with amplitudes of $\sim1$ mag 
around BJD 2456765 and 2456767 (figure \ref{fig:lc}).
Such an oscillation phase of this object, which is characteristic of AM CVn-type objects 
(cf. \cite{ram11amcvn,lev11j0719,lev15amcvn}), 
was also observed in March 2012 \citep{Pdot5}.
The dips in AM CVn-type DNe are different from those in hydrogen-rich CVs (such as in WZ Sge-type DNe) 
in that they occur already early during the superoutburst plateau (cf. \cite{kat04v803cen,nog04v406hya}).
In WZ Sge-type DNe, dips or oscillations occur after the long superoutburst \citep{kat15wzsge}.
In helium accretion disks, it may be difficult to maintain the hot state during superoutburst.

\subsection{Orbital Period of CR Boo}\label{sec:orb}
The well-known orbital period of CR Boo is 0.0170290(6) d \citep{pro97crboo}.
This period was estimated by using the humps in quiescence.
Humps in quiescence are basically caused by the hot spot in binary systems. 
The periods of such humps are thus regarded as the same as the orbital period.
However, in binaries with an extreme high outburst frequency such as ER UMa-type objects,
the binary may always have an eccentric disk and may show ``permanent superhumps'' \citep{pat99SH}.
Since the CR Boo is an ER UMa-like object,
the humps in quiescence may be not orbital ones.
Note that the period of humps in quiescence in 2015
is consistent with the previously measured one.
The orbital period by time-resolved spectroscopy is needed to be confirmed.

\subsection{Estimation of the Mass Ratio from Stage A Superhumps}
According to the previous work, 
the mass ratio of CR Boo was estimated to be $0.085 \pm 0.045$ \citep{roe07amcvndistance}.

This value is given by the empirical formula with the maximum error of 50$\%$.
By using the new method derived by \citet{kat13qfromstageA}, 
we could estimate a mass ratio more accurately.
As mentioned in section \ref{sec:int}, the period of the stage A superhumps appears to reflect
the dynamical precession rate of the disk at $r_{\rm{3:1}}$.
Thus, the mass ratios can be measured by using the theoretical equation of the dynamical precession rate
which was derived by \citep{hir90SHexcess}.
This value is deflected only by the errors of $P_{\rm orb}$ and $P_{\rm SH}$ of stage A, so it has small error
(for more detail, see \cite{kat13qfromstageA}).

The fractional superhump excess ($\varepsilon^{*} \equiv 1-P_{\rm orb}/P_{\rm SH}$) was 
0.0362(13) in 2014 and 0.0319(11) in 2015,
and we obtained mass ratios of 0.101(4) and 0.087(3), respectively.
The former value 0.101(4) was adopted 
as the final mass ratio because of the better $O-C$ diagram as described in section \ref{sec:superhump}.

\subsection{Mass-ratio and Implications on the Evolutionary Status}
Based on equations (\ref{eq:fulldege}) and (\ref{eq:semidege}),
we can draw the theoretical evolutionary tracks,
assuming that the secondary star is semi-degenerate or fully-degenerate \citep{tsu97amcvn}.
If we know a sufficient number of binary mass ratios,
we can compare the observation with the evolutionary population synthesis of AM CVn-type objects.

\begin{figure}
  \begin{center}
    \FigureFile(80mm,50mm){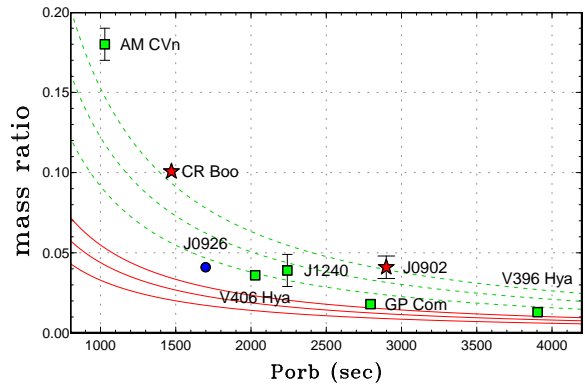}
  \end{center}
  \caption{Relation between $q$ and $P_{\rm{orb}}$ with various primary masses.
The green dashed curve indicates semi-degenerate secondaries,
and the red solid curve indicates fully-degenerate secondaries.
From top to bottom, three lines represent the $q$-$P_{\rm{orb}}$ relation, assuming 
$M_1 =$ 0.60, 0.75, and 1.00 $M_{\odot}$, respectively.
The filled stars represent the measurements from stage A superhump 
(J0902 (SDSS J090221.35+381941.9): \cite{kat14j0902}).
The filled squares represent the measurements from Doppler tomography \citep{DopplerTomography}
(AM CVn: \cite{roe06amcvn}, V406 Hya: \cite{roe06v406hya}, J1240 (SDSS J124058.03−015919): \cite{roe05j1240},
GP Com: \cite{mar99gpcom}, V396 Hya \cite{sol10amcvnreview}).
The filled circle represents the measurement from eclipse observations \citep{cop11j0926}
(J0926 (SDSS J092620.42+034542.3): \cite{cop11j0926}).
}
  \label{fig:dege_all}
\end{figure}

We plotted the evolutionary tracks with objects whose orbital period and
mass ratio are known in figure \ref{fig:dege_all}.
The green dashed curves indicate semi-degenerate secondaries,
and the red solid curves indicate fully-degenerate secondaries, respectively.
From top to bottom, three lines represent the $q$-$P_{\rm{orb}}$ relation, 
assuming $M_1 =$ 0.60, 0.75, and 1.00 $M_{\odot}$, respectively.
Basically, assuming that the same orbital period and primary mass, 
the semi-degenerate secondary has a higher mass-transfer rate than the fully-degenerate one.
Because the mass of the semi-degenerate secondary is needed to be heavier due to the lower density,
such a binary emits stronger GW
(see figures 4,5 in \cite{nel01amcvnpopulationsynthesis}).
Figure \ref{fig:dege_all} shows the secondary of CR Boo was semi-degenerate or almost non-degenerate.
We could confirm that the estimated mass ratio falls within the expected range.
There remain, however, possibilities that the location in the non-degenerate range may be incorrect:
(1) The primary mass is smaller than the typical DNe.
(2) The stage A superhump period was overestimated due to lacking data.
(3) The known orbital period is incorrect (see section \ref{sec:orb}).

All outbursting objects (CR Boo, V406 Hya, SDSS J092620.42+034542.3, 
SDSS J124058.03−015919, and SDSS J090221.35+381941.9)
are in the semi-degenerate range (see figure \ref{fig:dege_all}: 
note that the methods of estimating $q$ are not homogeneous).
The high occurrence of objects in the semi-degenerate
region may have been a result of a selection bias,
since such objects are expected to have higher
mass-transfer rates and they are more likely detected
due to the high occurrence of outbursts.

\subsection{Pressure Effect in Helium Disks}
\citet{kat14j0902} suggested that we can estimate the contribution of 
the pressure effect by the difference between $\varepsilon^*$ of stage A and B.
The aspidal precession frequency $\nu_{\rm pr}$ is represented by
the following equation \citep{lub92SH}: $\nu_{\rm pr} = \nu_{\rm dyn} + \nu_{\rm pressure} + \nu_{\rm stress}$,
where  $\nu_{\rm dyn}$, $\nu_{\rm pressure}$, and $\nu_{\rm stress}$ are 
disk precession frequencies due to the dynamical force of the secondary,
the pressure effect, and the minor wave-wave interaction, respectively.
According to \citet{kat13qfromstageA}, stage A depends only on the first term
and stage B depends on the first term and second term if we can ignore $\nu_{\rm stress}$.
Therefore $\varepsilon^*$(stage A)$-\varepsilon^*$(stage B) corresponds to $\nu_{\rm pressure}/\nu_{\rm orb}$.

$\varepsilon^*$(stage A)$-\varepsilon^*$(stage B) of CR Boo are
0.025(2) in 2014 and 0.020(4) in 2015.
These values are larger than those of hydrogen-rich CVs (0.010--0.015 \citet{Pdot})
and are consistent with the suggestion by \citet{pea07amcvnSH}
that the pressure effect in helium-rich disks appears to be higher than in hydrogen-rich disks
because of the higher ionization temperature.

\section{Summary}
We report on photometric observations of two superoutbursts 
of CR Boo in 2014 April--May and 2015 May--June.
we detected growing (stage A) superhumps and 
a precursor outburst for the second time in AM CVn-type objects,
and for the first time in typical outbursting AM CVn-type objects,
whose orbital periods are 1300--2500 s.
Although the outburst phenomena in AM CVn-type objects are widely known,
the observations have been insufficient yet.
Our observations therefore support that we can discuss 
AM CVn-type objects in analogy to hydrogen-rich CVs.
However, during the plateau phase, we also detected 
characteristic oscillations of AM CVn-type objects, called dips.
WZ Sge-type DNe, hydrogen-rich CVs, also show similar oscillations,
nevertheless the dips or oscillations occur after a long superoutburst, unlike in AM CVn-type objects.
This difference may be caused by the difficulty of maintaining the hot state.

As a result of our analyses by the PDM method and the $O-C$ diagrams,
we could estimate the mean superhump periods.
The periods of stage A, B, and C in 2014 are
0.017669(24), 0.017241(18) and 0.017218(5) d, respectively.
Moreover, the periods of stage A and B in 2015 are
0.017591(20) and 0.017237(48), respectively.
The period derivatives of stage B $P_{\rm{dot}}$ are
$1.01(29) \times 10^{-5}$ in 2014 and $1.76(37) \times 10^{-5}$ in 2015.
Assuming that the stage B continued before the dips,
the re-estimated $P_{\rm{dot}}$ in 2014 is $1.76(12) \times 10^{-5}$.
This value and $P_{\rm{dot}}$ in 2015 are consistent with 
the $P_{\rm{dot}}$ in 2012 of $2.0(0.2) \times 10^{-5}$.

By the new method using the period of stage A superhumps,
we obtained the mass ratios of CR Boo, $q=$0.101(4) in 2014 and 0.087(3) in 2015, respectively.
We adopted the former value because of the more clear $O-C$ diagram as described in section \ref{sec:superhump}.
This mass ratio was confirmed to lie on the theoretical evolutionary path.
To verify the theory, more measurements of the mass ratios of AM CVn-type objects are needed.
It is expected that this method will be applied to many objects which show superoutbursts
regardless of whether the objects are hydrogen-rich or helium-rich.

\section*{Acknowledgments}
This work was supported by the Grant-in-Aid Initiative for 
High-Dimensional Data-Driven Science through Deepening of Sparse Modeling (25120007)
from the Ministry of Education, Culture, Sports, 
Science and Technology (MEXT) of Japan. 
We acknowledge with thanks the variable-star observations 
by observers worldwide and used in this research.

\begin{table*}
\caption{Log of observations of CR Boo in 2014 April--May.}
\begin{center}
\scriptsize
\begin{tabular}{cccccccccccccc}
\hline
Start\commenta & End\commenta & mag\commentb & error\commentc & $N$\commentd & obs\commente & sys\commentf & Start\commenta & End\commenta & mag\commentb & error\commentc & $N$\commentd & obs\commente & sys\commentf \\
\hline
54.1681 & 54.2967 & 1.4740 & 0.0100 & 165 & OKU & C & 64.0963 & 64.1686 & 0.7480 & 0.0090 & 23 & OKU & C \\
54.1700 & 54.3069 & 15.6650 & 0.0090 & 283 & Ioh & C & 64.7275 & 64.8408 & 14.9930 & 0.0220 & 52 & HaC & CV \\
55.1924 & 55.2802 & 2.3700 & 0.0070 & 77 & OKU & C & 64.7892 & 64.8750 & 14.7920 & 0.0030 & 173 & OkC & CV \\
55.7550 & 55.8673 & 17.2140 & 0.0200 & 43 & HaC & CV & 65.3605 & 65.4357 & 14.7690 & 0.0020 & 195 & Kai & C \\
56.2237 & 56.3082 & 2.8860 & 0.0130 & 74 & OKU & C & 65.6687 & 65.8750 & 14.4710 & 0.0020 & 413 & OkC & CV \\
56.3065 & 56.3710 & 6.3280 & 0.0150 & 88 & CRI & C & 66.5911 & 66.8764 & 15.0500 & 0.0080 & 1000 & LCO & CV \\
56.7574 & 56.8646 & 17.2820 & 0.0190 & 36 & HaC & CV & 66.7220 & 66.8749 & 15.2100 & 0.0070 & 339 & HaC & CV \\
57.0446 & 57.1938 & 4.3200 & 0.0080 & 359 & KU1 & C & 67.1764 & 67.1859 & 0.5660 & 0.0050 & 14 & OKU & C \\
57.1928 & 57.2818 & 1.9790 & 0.0070 & 76 & OKU & C & 67.3282 & 67.4310 & 2.7110 & 0.0020 & 121 & DPV & C \\
57.3192 & 57.3727 & 5.0500 & 0.0120 & 73 & CRI & C & 67.7191 & 67.8326 & 14.8960 & 0.0050 & 52 & HaC & CV \\
57.3939 & 57.6675 & 15.4880 & 0.0120 & 713 & Kai & C & 68.3276 & 68.5555 & 3.0730 & 0.0030 & 260 & Nov & C \\
57.7470 & 57.8625 & 14.6610 & 0.0030 & 52 & HaC & CV & 68.7163 & 68.8288 & 14.6760 & 0.0050 & 42 & HaC & CV \\
57.9936 & 58.2124 & 14.7350 & 0.0070 & 477 & Kis & C & 69.2196 & 69.2735 & 0.4550 & 0.0050 & 50 & OKU & C \\
58.0558 & 58.3436 & 14.0390 & 0.0050 & 656 & Mdy & C & 69.7137 & 69.8253 & 14.8100 & 0.0030 & 42 & HaC & CV \\
58.1756 & 58.2641 & -0.2750 & 0.0030 & 213 & OKU & C & 69.7385 & 69.8607 & 14.7300 & 0.0020 & 248 & OkC & CV \\
58.1996 & 58.3214 & 1.9020 & 0.0030 & 267 & KU1 & C & 70.1754 & 70.2336 & 0.4130 & 0.0020 & 54 & OKU & C \\
58.2946 & 58.3719 & 3.4020 & 0.0040 & 196 & CRI & C & 70.7109 & 70.7895 & 14.7150 & 0.0030 & 28 & HaC & CV \\
58.3278 & 58.6227 & 2.0000 & 0.0020 & 579 & Mas & C & 70.7760 & 70.8544 & 14.6670 & 0.0020 & 158 & OkC & CV \\
58.3346 & 58.5850 & 3.3090 & 0.0020 & 651 & CRI & C & 71.1723 & 71.2848 & 0.3800 & 0.0020 & 127 & OKU & C \\
58.3985 & 58.6691 & 14.0720 & 0.0020 & 1314 & Kai & C & 72.0138 & 72.1165 & 15.2100 & 0.0030 & 229 & Kis & C \\
58.7441 & 58.8598 & 14.0730 & 0.0050 & 55 & HaC & CV & 72.0258 & 72.1514 & 14.6300 & 0.0010 & 253 & Mdy & C \\
58.7627 & 58.8750 & 14.0780 & 0.0030 & 211 & OkC & CV & 72.8358 & 72.8542 & 14.7710 & 0.0040 & 38 & OkC & CV \\
59.0349 & 59.2980 & 13.9560 & 0.0020 & 521 & Ioh & C & 72.9834 & 73.0712 & 15.3350 & 0.0070 & 198 & Kis & C \\
59.0736 & 59.2187 & 1.9450 & 0.0020 & 371 & KU1 & C & 73.1046 & 73.2750 & 14.7160 & 0.0040 & 324 & Ioh & C \\
59.1364 & 59.2625 & -0.1720 & 0.0020 & 297 & OKU & C & 73.1143 & 73.2392 & 0.4660 & 0.0020 & 171 & OKU & C \\
59.3580 & 59.4768 & 2.1270 & 0.0020 & 236 & Mas & C & 73.1531 & 73.2165 & 2.6860 & 0.0080 & 142 & KU1 & C \\
59.4124 & 59.5071 & 14.1520 & 0.0020 & 256 & Kai & C & 73.7512 & 73.8498 & 14.8390 & 0.0020 & 200 & OkC & CV \\
59.7413 & 59.8569 & 14.2090 & 0.0040 & 56 & HaC & CV & 73.9802 & 74.0260 & 15.3850 & 0.0040 & 103 & Kis & C \\
60.0589 & 60.2530 & 14.1050 & 0.0040 & 206 & Ioh & C & 73.9960 & 74.1623 & 14.7470 & 0.0030 & 321 & Ioh & C \\
60.1337 & 60.2739 & 14.7650 & 0.0050 & 146 & Kis & C & 74.0361 & 74.1370 & 2.6890 & 0.0030 & 260 & KU1 & C \\
60.1680 & 60.2072 & -0.0310 & 0.0070 & 35 & OKU & C & 74.1611 & 74.2474 & 0.5920 & 0.0040 & 79 & OKU & C \\
60.3096 & 60.4041 & 2.6880 & 0.0100 & 66 & Nov & C & 74.7513 & 74.8542 & 14.8990 & 0.0020 & 207 & OkC & CV \\
60.3930 & 60.8542 & 14.4370 & 0.0090 & 116 & HaC & CV & 74.9662 & 75.1626 & 14.8880 & 0.0020 & 380 & Ioh & C \\
60.7138 & 60.8864 & 14.2750 & 0.0020 & 350 & OkC & CV & 75.3170 & 75.4251 & 2.8840 & 0.0030 & 136 & DPV & C \\
62.0043 & 62.1320 & 14.9800 & 0.0030 & 284 & Kis & C & 75.7166 & 75.8543 & 16.3490 & 0.0060 & 274 & OkC & CV \\
62.0579 & 62.1220 & 2.3100 & 0.0040 & 144 & KU1 & C & 76.7482 & 76.8539 & 16.9250 & 0.0050 & 125 & OkC & CV \\
62.1143 & 62.2780 & 0.1150 & 0.0030 & 184 & OKU & C & 77.4193 & 77.5626 & 1.4870 & 0.0060 & 146 & CRI & C \\
62.7587 & 62.8731 & 14.5580 & 0.0020 & 231 & OkC & CV & 77.7527 & 77.8545 & 17.0250 & 0.0070 & 97 & OkC & CV \\
63.0037 & 63.0701 & 2.5530 & 0.0110 & 101 & KU1 & C & 79.3313 & 79.4719 & 0.2680 & 0.0090 & 172 & CRI & C \\
63.0319 & 63.1676 & 15.3350 & 0.0100 & 294 & Kis & C & 80.0446 & 80.0799 & 4.8470 & 0.0220 & 81 & KU1 & C \\
\hline
\multicolumn{14}{l}{\commenta JD$-$2456700.} \\
\multicolumn{14}{l}{\commentb Mean magnitude.} \\
\multicolumn{14}{l}{\commentc 1 $\sigma$ of the mean magnitude.} \\
\multicolumn{14}{l}{\commentd Number of observations.} \\
\multicolumn{14}{l}{\parbox{240pt}{\commente Observer's code: Kis (S. Kiyota), Ioh (H. Itoh), OKU (Osaka Kyoiku U. team), 
 HaC (F. J. Hambsch), CRI (Crimean Astrophys. Obs.), KU1 (Kyoto U. team), 
 Kai (K. Kasai), Mdy (Y. Maeda), Mas (G. Masi), OkC ( A. Oksanen), 
LCO (C. Littlefield), Nov (R. Novak), DPV (P. Dubovsky)}}\\
\multicolumn{7}{l}{\commentf Filter.  ``C'' means no filter (clear).} \\
\end{tabular}
\end{center}
\end{table*}

\begin{table*}
\caption{Log of observations of CR Boo in 2015 May--June.}\label{tab:log_crboo2015}
\begin{center}
\scriptsize
\begin{tabular}{ccccccc}
\hline
Start\commenta & End\commenta & mag\commentb & error\commentc & $N$\commentd & obs\commente & sys\commentf \\
\hline
68.9675 & 69.2310 & 3.656 & 0.009 & 451 & KU1 & C \\
69.3674 & 69.4176 & 16.924 & 0.031 & 23 & COO & CV \\
69.6436 & 69.6979 & 16.727 & 0.027 & 44 & COO & CV \\
69.6906 & 69.8657 & 16.627 & 0.009 & 154 & COO & CV \\
69.9616 & 70.2212 & 3.180 & 0.008 & 572 & KU1 & C \\
70.0089 & 70.0122 & 4.316 & 0.009 & 10 & OKU & C \\
70.3539 & 70.5471 & 16.751 & 0.009 & 457 & Kai & C \\
70.4783 & 70.7267 & 17.073 & 0.036 & 78 & HaC & CV \\
71.3486 & 71.4785 & 4.172 & 0.014 & 138 & DPV & C \\
71.4765 & 71.7232 & 17.398 & 0.066 & 64 & HaC & CV \\
71.6371 & 71.7719 & 16.944 & 0.030 & 87 & COO & CV \\
71.9593 & 72.2293 & 3.084 & 0.019 & 501 & KU1 & C \\
71.9597 & 71.9630 & 5.032 & 0.040 & 10 & OKU & C \\
72.3420 & 72.4876 & 14.857 & 0.020 & 143 & Kai & C \\
72.4770 & 72.7205 & 14.497 & 0.011 & 55 & HaC & CV \\
72.5776 & 72.7845 & 0.504 & 0.003 & 219 & DKS & C \\
72.7101 & 72.9055 & 14.112 & 0.008 & 151 & COO & CV \\
73.9596 & 74.2278 & 0.754 & 0.005 & 387 & KU1 & C \\
74.0005 & 74.0801 & 13.833 & 0.007 & 67 & COO & V \\
74.0044 & 74.2081 & 1.837 & 0.002 & 544 & OKU & C \\
74.4760 & 74.7158 & 14.195 & 0.004 & 81 & HaC & CV \\
75.3507 & 75.4986 & 0.770 & 0.003 & 192 & DPV & C \\
75.6881 & 75.7365 & 14.291 & 0.006 & 171 & COO & CV \\
76.3280 & 76.4669 & 0.903 & 0.004 & 168 & CRI & C \\
76.3420 & 76.4845 & 0.894 & 0.003 & 185 & DPV & C \\
76.4656 & 76.7113 & 14.435 & 0.003 & 86 & HaC & CV \\
76.4703 & 76.5224 & 14.021 & 0.006 & 23 & COO & CV \\
76.6661 & 76.7361 & 0.221 & 0.003 & 52 & COO & CV \\
76.7667 & 76.8180 & 14.181 & 0.004 & 44 & COO & CV \\
77.0188 & 77.1308 & 1.055 & 0.002 & 294 & KU1 & C \\
80.2850 & 80.4749 & 1.276 & 0.002 & 241 & CRI & C \\
81.3053 & 81.4793 & 1.147 & 0.006 & 221 & CRI & C \\
82.3154 & 82.4890 & 1.525 & 0.018 & 83 & CRI & C \\
83.3337 & 83.4908 & 1.231 & 0.006 & 138 & CRI & C \\
\hline
\multicolumn{7}{l}{\commenta JD$-$2457100.} \\
\multicolumn{7}{l}{\commentb Mean magnitude.} \\
\multicolumn{7}{l}{\commentc 1 $\sigma$ of the mean magnitude.} \\
\multicolumn{7}{l}{\commentd Number of observations.} \\
\multicolumn{7}{l}{\parbox{240pt}{\commente Observer's code: L. M. Cook (COO), OKU (Osaka Kyoiku U.), 
 HaC (F. J. Hambsch), CRI (Crimean Astrophys. Obs.), KU1 (Kyoto University), 
 Kai (K. Kasai), DPV (P. Dubovsky), S. Dvorak (DKS)}}\\
\multicolumn{7}{l}{\commentf Filter.  ``C'' means no filter (clear).} \\
\end{tabular}
\end{center}
\end{table*}

\begin{table*}
\caption{Times of superhump maxima in CR Boo 2014 April--May.}
\begin{center}
\scriptsize
\begin{tabular}{cccccccccc}
\hline
$E$ & maximum time\commenta & error\commentb & $O-C$\commentc & $N$\commentd & $E$ & maximum time\commenta & error\commentb & $O-C$\commentc & $N$\commentd \\
\hline
0 & 56758.00770 & 0.00080 & -0.00340 & 30 & 236 & 56762.08140 & 0.00080 & 0.00130 & 62 \\
1 & 56758.02890 & 0.00180 & 0.00060 & 30 & 238 & 56762.11490 & 0.00040 & 0.00040 & 60 \\
4 & 56758.07860 & 0.00130 & -0.00150 & 60 & 240 & 56762.15170 & 0.00070 & 0.00260 & 18 \\
5 & 56758.09640 & 0.00090 & -0.00090 & 64 & 244 & 56762.21800 & 0.00060 & -0.00010 & 19 \\
6 & 56758.11400 & 0.00060 & -0.00060 & 62 & 276 & 56762.77190 & 0.00030 & 0.00220 & 28 \\
7 & 56758.13070 & 0.00100 & -0.00110 & 66 & 277 & 56762.78940 & 0.00030 & 0.00240 & 28 \\
8 & 56758.14890 & 0.00110 & -0.00010 & 63 & 278 & 56762.80620 & 0.00030 & 0.00200 & 27 \\
9 & 56758.16700 & 0.00040 & 0.00070 & 53 & 280 & 56762.84050 & 0.00030 & 0.00190 & 28 \\
10 & 56758.18560 & 0.00040 & 0.00210 & 88 & 281 & 56762.85850 & 0.00030 & 0.00260 & 27 \\
11 & 56758.20170 & 0.00040 & 0.00090 & 176 & 445 & 56765.67670 & 0.00050 & 0.01040 & 19 \\
12 & 56758.21850 & 0.00030 & 0.00050 & 142 & 446 & 56765.69460 & 0.00030 & 0.01110 & 21 \\
13 & 56758.23610 & 0.00020 & 0.00090 & 99 & 447 & 56765.71140 & 0.00030 & 0.01070 & 28 \\
14 & 56758.25330 & 0.00020 & 0.00080 & 87 & 448 & 56765.72850 & 0.00020 & 0.01050 & 28 \\
15 & 56758.26910 & 0.00020 & -0.00070 & 66 & 449 & 56765.74590 & 0.00020 & 0.01060 & 28 \\
16 & 56758.28740 & 0.00020 & 0.00040 & 59 & 450 & 56765.76350 & 0.00030 & 0.01100 & 28 \\
17 & 56758.30420 & 0.00020 & 0.00000 & 95 & 451 & 56765.78100 & 0.00020 & 0.01120 & 28 \\
18 & 56758.32080 & 0.00020 & -0.00070 & 101 & 452 & 56765.79800 & 0.00020 & 0.01100 & 28 \\
19 & 56758.33790 & 0.00010 & -0.00080 & 101 & 453 & 56765.81540 & 0.00030 & 0.01120 & 28 \\
20 & 56758.35540 & 0.00010 & -0.00050 & 99 & 454 & 56765.83260 & 0.00020 & 0.01110 & 27 \\
21 & 56758.37250 & 0.00020 & -0.00070 & 93 & 455 & 56765.84970 & 0.00030 & 0.01100 & 28 \\
22 & 56758.39000 & 0.00010 & -0.00040 & 64 & 456 & 56765.86760 & 0.00040 & 0.01170 & 29 \\
23 & 56758.40740 & 0.00020 & -0.00030 & 116 & 680 & 56769.74750 & 0.00030 & 0.02950 & 30 \\
24 & 56758.42490 & 0.00020 & 0.00000 & 135 & 681 & 56769.76300 & 0.00030 & 0.02780 & 33 \\
26 & 56758.45970 & 0.00010 & 0.00030 & 134 & 682 & 56769.78220 & 0.00090 & 0.02970 & 32 \\
27 & 56758.47650 & 0.00010 & -0.00020 & 133 & 683 & 56769.79990 & 0.00040 & 0.03010 & 33 \\
28 & 56758.49380 & 0.00010 & -0.00010 & 132 & 684 & 56769.81590 & 0.00030 & 0.02890 & 34 \\
29 & 56758.51090 & 0.00010 & -0.00020 & 133 & 685 & 56769.83260 & 0.00060 & 0.02840 & 31 \\
31 & 56758.54540 & 0.00010 & -0.00010 & 132 & 686 & 56769.85160 & 0.00070 & 0.03020 & 26 \\
32 & 56758.56220 & 0.00030 & -0.00070 & 100 & 740 & 56770.78350 & 0.00050 & 0.03100 & 25 \\
33 & 56758.58000 & 0.00020 & -0.00010 & 131 & 741 & 56770.80080 & 0.00040 & 0.03100 & 29 \\
34 & 56758.59740 & 0.00020 & 0.00010 & 94 & 742 & 56770.81900 & 0.00060 & 0.03210 & 28 \\
35 & 56758.61460 & 0.00020 & 0.00000 & 94 & 743 & 56770.83540 & 0.00060 & 0.03120 & 28 \\
36 & 56758.63180 & 0.00020 & 0.00000 & 73 & 744 & 56770.85280 & 0.00030 & 0.03140 & 25 \\
37 & 56758.64880 & 0.00020 & -0.00020 & 66 & 764 & 56771.19880 & 0.00070 & 0.03250 & 19 \\
38 & 56758.66630 & 0.00040 & 0.00010 & 64 & 765 & 56771.21570 & 0.00090 & 0.03220 & 19 \\
44 & 56758.76820 & 0.00090 & -0.00150 & 15 & 766 & 56771.23220 & 0.00040 & 0.03150 & 19 \\
45 & 56758.78740 & 0.00040 & 0.00040 & 28 & 817 & 56772.11370 & 0.00320 & 0.03360 & 55 \\
46 & 56758.80370 & 0.00020 & -0.00050 & 35 & 913 & 56773.76330 & 0.00040 & 0.02810 & 28 \\
47 & 56758.82110 & 0.00020 & -0.00040 & 34 & 914 & 56773.78120 & 0.00060 & 0.02870 & 28 \\
48 & 56758.83830 & 0.00020 & -0.00040 & 36 & 915 & 56773.79770 & 0.00080 & 0.02800 & 27 \\
49 & 56758.85570 & 0.00030 & -0.00030 & 36 & 916 & 56773.81440 & 0.00070 & 0.02740 & 28 \\
50 & 56758.87260 & 0.00020 & -0.00060 & 28 & 918 & 56773.84930 & 0.00050 & 0.02780 & 27 \\
60 & 56759.04470 & 0.00050 & -0.00090 & 20 & 971 & 56774.76160 & 0.00020 & 0.02630 & 24 \\
61 & 56759.06210 & 0.00060 & -0.00070 & 26 & 972 & 56774.79500 & 0.00030 & 0.02530 & 28 \\
62 & 56759.07950 & 0.00030 & -0.00060 & 49 & 973 & 56774.81210 & 0.00040 & 0.02510 & 27 \\
63 & 56759.09670 & 0.00030 & -0.00070 & 62 & 975 & 56774.82980 & 0.00060 & 0.02560 & 28 \\
64 & 56759.11440 & 0.00040 & -0.00020 & 64 & 975 & 56774.84740 & 0.00060 & 0.02590 & 28 \\
65 & 56759.13100 & 0.00030 & -0.00080 & 64 & 987 & 56775.05000 & 0.00060 & 0.02160 & 29 \\
67 & 56759.16500 & 0.00020 & -0.00130 & 97 & 988 & 56775.06920 & 0.00130 & 0.02360 & 29 \\
68 & 56759.18300 & 0.00020 & -0.00050 & 98 & 990 & 56775.08900 & 0.00110 & 0.02610 & 26 \\
69 & 56759.19960 & 0.00020 & -0.00120 & 93 & 1006 & 56775.37930 & 0.00110 & 0.02330 & 18 \\
70 & 56759.21750 & 0.00020 & -0.00050 & 94 & 1007 & 56775.39820 & 0.00140 & 0.02500 & 18 \\
71 & 56759.23450 & 0.00020 & -0.00080 & 56 & 1008 & 56775.41490 & 0.00100 & 0.02450 & 16 \\
72 & 56759.25170 & 0.00020 & -0.00080 & 55 & 1027 & 56775.74140 & 0.00030 & 0.02340 & 28 \\
73 & 56759.26900 & 0.00020 & -0.00070 & 39 & 1028 & 56775.75860 & 0.00020 & 0.02340 & 28 \\
74 & 56759.28600 & 0.00030 & -0.00100 & 30 & 1029 & 56775.77610 & 0.00030 & 0.02360 & 28 \\
75 & 56759.30310 & 0.00040 & -0.00110 & 15 & 1030 & 56775.79320 & 0.00070 & 0.02340 & 28 \\
79 & 56759.37230 & 0.00030 & -0.00090 & 27 & 1031 & 56775.80980 & 0.00070 & 0.02280 & 23 \\
80 & 56759.38960 & 0.00020 & -0.00090 & 28 & 1032 & 56775.82740 & 0.00060 & 0.02320 & 28 \\
81 & 56759.40670 & 0.00020 & -0.00100 & 28 & 1033 & 56775.84450 & 0.00120 & 0.02310 & 27 \\
82 & 56759.42380 & 0.00020 & -0.00110 & 62 & 1088 & 56776.79230 & 0.00040 & 0.02250 & 17 \\
83 & 56759.44140 & 0.00020 & -0.00070 & 65 & 1090 & 56776.82640 & 0.00080 & 0.02210 & 16 \\
84 & 56759.45870 & 0.00020 & -0.00070 & 65 & 1091 & 56776.84390 & 0.00050 & 0.02250 & 16 \\
86 & 56759.49290 & 0.00030 & -0.00090 & 37 & 1125 & 56777.43170 & 0.00040 & 0.02400 & 13 \\
87 & 56759.51030 & 0.00040 & -0.00080 & 25 & 1126 & 56777.44590 & 0.00050 & 0.02100 & 17 \\
121 & 56760.09600 & 0.00060 & -0.00140 & 28 & 1127 & 56777.46480 & 0.00080 & 0.02260 & 17 \\
124 & 56760.14640 & 0.00080 & -0.00260 & 26 & 1131 & 56777.53360 & 0.00070 & 0.02240 & 17 \\
128 & 56760.21600 & 0.00050 & -0.00200 & 34 & 1132 & 56777.55190 & 0.00120 & 0.02360 & 17 \\
129 & 56760.23360 & 0.00050 & -0.00170 & 33 & 1133 & 56777.56670 & 0.00040 & 0.02100 & 10 \\
130 & 56760.25130 & 0.00040 & -0.00120 & 39 & 1145 & 56777.77420 & 0.00060 & 0.02170 & 16 \\
158 & 56760.73420 & 0.00020 & -0.00110 & 29 & 1146 & 56777.79210 & 0.00040 & 0.02230 & 16 \\
159 & 56760.75110 & 0.00020 & -0.00140 & 33 & 1147 & 56777.80920 & 0.00060 & 0.02220 & 15 \\
160 & 56760.76880 & 0.00020 & -0.00100 & 34 & 1149 & 56777.84280 & 0.00030 & 0.02130 & 17 \\
161 & 56760.78600 & 0.00020 & -0.00100 & 34 & 1237 & 56779.35590 & 0.00060 & 0.01720 & 17 \\
162 & 56760.80300 & 0.00030 & -0.00130 & 34 & 1238 & 56779.37600 & 0.00050 & 0.02000 & 17 \\
163 & 56760.82090 & 0.00040 & -0.00060 & 33 & 1239 & 56779.39110 & 0.00030 & 0.01790 & 17 \\
164 & 56760.83720 & 0.00040 & -0.00150 & 36 & 1240 & 56779.40870 & 0.00030 & 0.01830 & 17 \\
165 & 56760.85490 & 0.00040 & -0.00110 & 34 & 1241 & 56779.42560 & 0.00030 & 0.01790 & 17 \\
166 & 56760.87170 & 0.00030 & -0.00140 & 27 & 1242 & 56779.44320 & 0.00030 & 0.01830 & 17 \\
167 & 56760.88960 & 0.00030 & -0.00080 & 19 & 1243 & 56779.46040 & 0.00030 & 0.01820 & 17 \\
\hline
  \multicolumn{10}{l}{\commenta BJD$-$2400000.} \\
  \multicolumn{10}{l}{\commentb Unit: day} \\
  \multicolumn{10}{l}{\commentc  $C= 2456758.0111 + 0.0172414 E$.} \\
  \multicolumn{10}{l}{\commentd Number of points used to determine the maximum.} \\
\end{tabular}
\end{center}
\end{table*}

\begin{table*}
\caption{Times of superhump and orbital hump maxima in CR Boo 2015 May--June.}
\begin{center}
\scriptsize
\begin{tabular}{ccccc}
\hline
$E$ & maximum time\commenta & error\commentb & $O-C$\commentc & $N$\commentd\\
\hline
0 & 57168.99429 & 0.00178 & 0.02734 & 36 \\
1 & 57169.00862 & 0.00045 & 0.02442 & 36 \\
2 & 57169.02734 & 0.00049 & 0.02590 & 33 \\
8 & 57169.12903 & 0.00065 & 0.02415 & 33 \\
9 & 57169.14664 & 0.00046 & 0.02451 & 31 \\
10 & 57169.16237 & 0.00056 & 0.02300 & 33 \\
42 & 57169.70687 & 0.00054 & 0.01577 & 16 \\
43 & 57169.72404 & 0.00048 & 0.01570 & 11 \\
44 & 57169.74195 & 0.00047 & 0.01637 & 8 \\
45 & 57169.75781 & 0.00048 & 0.01499 & 13 \\
46 & 57169.77417 & 0.00029 & 0.01411 & 14 \\
47 & 57169.79235 & 0.00108 & 0.01505 & 10 \\
48 & 57169.80858 & 0.00048 & 0.01403 & 11 \\
49 & 57169.82618 & 0.00061 & 0.01439 & 15 \\
51 & 57169.86102 & 0.00043 & 0.01475 & 15 \\
63 & 57170.06509 & 0.00078 & 0.01193 & 36 \\
64 & 57170.08129 & 0.00101 & 0.01089 & 36 \\
67 & 57170.13212 & 0.00037 & 0.00999 & 35 \\
156 & 57171.64934 & 0.00054 & -0.00728 & 14 \\
185 & 57172.14455 & 0.00090 & -0.01207 & 18 \\
214 & 57172.63720 & 0.00097 & -0.01942 & 16 \\
220 & 57172.74279 & 0.00118 & -0.01728 & 22 \\
223 & 57172.79327 & 0.00051 & -0.01852 & 13 \\
290 & 57173.96716 & 0.00023 & 0.00020 & 28 \\
291 & 57173.98436 & 0.00026 & 0.00016 & 36 \\
293 & 57174.01894 & 0.00017 & 0.00026 & 57 \\
294 & 57174.03623 & 0.00016 & 0.00031 & 47 \\
295 & 57174.05340 & 0.00022 & 0.00023 & 50 \\
296 & 57174.07055 & 0.00023 & 0.00014 & 52 \\
297 & 57174.08743 & 0.00027 & -0.00022 & 41 \\
298 & 57174.10491 & 0.00014 & 0.00002 & 43 \\
299 & 57174.12196 & 0.00016 & -0.00017 & 73 \\
300 & 57174.13928 & 0.00018 & -0.00010 & 62 \\
301 & 57174.15662 & 0.00016 & -0.00000 & 72 \\
302 & 57174.17347 & 0.00047 & -0.00039 & 70 \\
303 & 57174.19084 & 0.00032 & -0.00026 & 43 \\
371 & 57175.36334 & 0.00017 & -0.00018 & 18 \\
372 & 57175.37968 & 0.00027 & -0.00107 & 18 \\
373 & 57175.39724 & 0.00026 & -0.00076 & 18 \\
374 & 57175.41414 & 0.00030 & -0.00110 & 17 \\
375 & 57175.43159 & 0.00022 & -0.00089 & 18 \\
376 & 57175.44888 & 0.00040 & -0.00084 & 18 \\
377 & 57175.46609 & 0.00026 & -0.00088 & 18 \\
378 & 57175.48312 & 0.00023 & -0.00108 & 18 \\
429 & 57176.36466 & 0.00039 & 0.00115 & 34 \\
431 & 57176.39703 & 0.00040 & -0.00097 & 35 \\
432 & 57176.41563 & 0.00032 & 0.00038 & 32 \\
433 & 57176.43177 & 0.00067 & -0.00072 & 36 \\
434 & 57176.44981 & 0.00037 & 0.00009 & 35 \\
437 & 57176.50347 & 0.00104 & 0.00202 & 11 \\
468 & 57177.03683 & 0.00030 & 0.00090 & 36 \\
469 & 57177.05359 & 0.00037 & 0.00042 & 37 \\
470 & 57177.07107 & 0.00069 & 0.00066 & 36 \\
471 & 57177.08749 & 0.00055 & -0.00017 & 36 \\
472 & 57177.10666 & 0.00029 & 0.00176 & 37 \\
473 & 57177.12330 & 0.00038 & 0.00117 & 37 \\
\hline
  \multicolumn{5}{l}{\commenta BJD$-$2400000.} \\
  \multicolumn{5}{l}{\commentb Unit: day} \\
  \multicolumn{5}{l}{\commentc $C= 2457168.99251 + 0.0172414 E$.} \\
  \multicolumn{5}{l}{\commentd Number of points used to determine the maximum.} \\
\end{tabular}
\end{center}
\end{table*}

\end{document}